\documentclass[twoside]{article}

\usepackage{fleqn,espcrc2}
\usepackage{epsfig}

\usepackage{graphicx}
\usepackage[figuresright]{rotating}

\newcommand{\AmS}{{\protect\the\textfont2
  A\kern-.1667em\lower.5ex\hbox{M}\kern-.125emS}}
\newcommand{\mnut}{M_{\nu_\tau}}

\title{
\begin{flushright}
 CLNS 00/1714
\end{flushright}
A Biased Review of Tau Neutrino Mass Limits
}

\author{J.~E. Duboscq\address[MCSD]{Wilson Laboratory, 
        Cornell University, \\ 
        Ithaca NY, 14853, USA \\
To appear in the proceedings of the 6th International Workshop
on Tau Lepton Physics, 18-21 September 2000, Victoria (Canada)}}
       
\begin{document}

\begin{abstract}
After a quick review of astrophysically relevant limits,
I present a summary of MeV scale tau neutrino mass limits derived from 
 accelerator based experiments. I argue that the current
 published limits appear to be too consistent, and that we therefore
 cannot conclude that the tau neutrino mass limit is as low
 as usually claimed. I provide motivational arguments calling into
 question the assumed statistical 
 properties of the usual maximum likelihood estimators, and 
 provide a prescription for deriving a more robust and 
 understandable mass limit.
\vspace{1pc}
\end{abstract}

\maketitle

\section{ Neutrinos and Cosmology  }
 All particle species, including neutrinos, must have been produced during 
the Big Bang. Should  neutrinos be stable and sufficiently massive, their 
combined gravitational attraction would be sufficient to collapse the 
Universe.  Our existence therefore allows one to infer properties of invisible 
but gravitationally interacting matter. This argument was originally advanced 
by Gershtein and Zel'dovich\cite{Zeldovich66} in 1966  and was used by Cowsik and McClelland\cite{Cowsik}
to devise a limit on 
the order of $100(\times h) eV$ for the sum of the masses of stable neutrino 
species,   
 where h is the scaled Hubble Constant. This  argument also 
supplies a lower bound on extremely heavy neutrinos at the GeV scale. 
 Details of these limits can change substantially if neutrinos decay. 
The change depends on the equation of state of its decay products - a 
massive neutrino decaying to massless particles will effect the evolution 
 of the universe 
differently than one with massive decay products. The effect also depends on 
the time scale of the neutrino decay relative to other time scales in 
the Universe, such as that of Big Bang nucleosynthesis.
The contribution of decaying neutrinos to supernovae energetics can
 also be used to derive limits on $\nu_\tau$ lifetime and mass \cite{Kolb}.
\begin{figure}[htb]
\centerline{
\epsfig{file=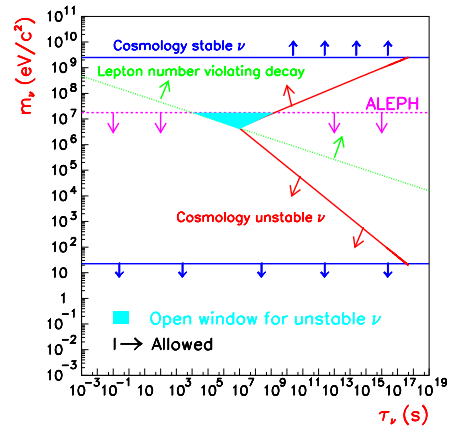,width=60mm, bbllx=1pt, 
 bblly=1pt, bburx=555pt, bbury=555pt ,clip=}
}
\caption{Some cosmological and particle physics limits on the $\nu_\tau$ mass and lifetime as per \cite{ALEPH98} }
\label{fig:cosmo}
\end{figure}
 Shown in Figure~\ref{fig:cosmo}
 are a set of  allowed and excluded regions of mass and lifetime, 
as supplied by ALEPH \cite{ALEPH98}.  In addition to these cosmological bounds, 
the figure also shows the limits imposed by the lack of observation of 
neutrino-less decays such as $\tau \to 3 { \mu}$ - these decays could have 
massive neutrinos inside of loops and would therefore be sensitive to 
neutrino properties. In addition, the stringent limit claimed by ALEPH\cite{ALEPH98}
 is also plotted 
in the figure. Note that on such a scale, the difference in mass limits 
from different experiments would be barely visible.
Of interest is 
that at the MeV mass scale, where accelerators are currently sensitive, 
many lines cross - this indicates that this could be a region rich 
with (mis)understanding, where many different approximations might break 
down.

  It would be interesting to review what effect recent indications 
of a non-zero cosmological constant might have on all these limits. The 
overlap of the Supernova  and Boomerang   data released 
this year indicate that the gravitational energy density of the Universe 
is to within a factor of a few equal to that used to derive the 
 Cowsik-McClelland 
limit and that the ensuing argument is still qualitatively valid. One 
can also recast the argument in terms of the stability of clusters, and 
this  still yields a comparable limit \cite{Turner00}.

\section{ Recent Results of  Interest }
The DONUT collaboration has recently announced that it has made the first
 observation of  tau  neutrino appearance. Although 
their results (reported elsewhere in these proceedings) do not tell us very 
much about the mass of this neutrino, they do tell us that it does  exist.

The KARMEN collaboration, which had previously reported an anomaly 
consistent with a  weakly interacting neutral particle  emitted 
in pion decays with a mass of 33.9 MeV \cite{KARMEN},  no longer sees 
this anomaly in 
a larger dataset. Their new results are reported elsewhere in these
 proceedings.

Possibly 
the most talked about recent results are those coming from the SuperKamiokande 
collaboration, also reported elsewhere in these proceedings. They report 
that their data is consistent with the mixing of muon type neutrinos into 
a different
 state, most probably  $\tau$ neutrinos. Should these results 
hold up, they are a strong argument against a massive neutrino in the 
allowed MeV mass range. Unfortunately, SuperKamiokande has yet to take 
into account the possibility that neutrinos are massive and can decay. 
Given that all other known particles that oscillate also decay, one might 
legitimately ask whether this is also the case in the neutrino sector. 
One could hope that low mass, long lived electron type neutrinos, medium 
lived muon neutrinos and short lived massive tau neutrinos might help 
explain the solar and atmospheric neutrino puzzles. Until such possibilities 
are ruled out it might be premature to assume that mixing has been proven 
and that the MeV mass scale for the tau neutrino is no longer
 interesting. 
 Simply showing consistency with a given model does not necessarily rule 
out other models.

\section{Accelerator Based Non-Endpoint Limits}
\begin{figure}[htb]
\centerline{
\epsfig{file=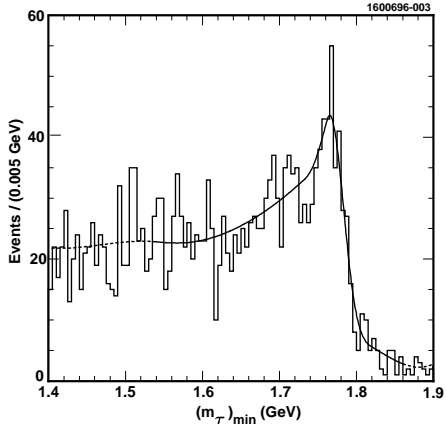,width=60mm,clip=}
}
\caption{Two Cone Mass for $m_\tau^2-m_\nu^2$ determination from \cite{CLEO97} - the location of the sharp drop determines the limit. }
\label{fig:twocone}
\end{figure}

At electron-positron colliders, it is possible to produce 
mono-energetic tau pairs. When these decay to hadrons and exactly one 
neutrino each, the tau flight direction on each side must have been along 
a cone around the direction of the hadronic decay products. The size of 
this cone depends on the tau and neutrino mass. Requiring that the 
common $\tau$ flight direction is on the overlap of the two cones
allows one to measure 
 $m_\tau^2-m_\nu^2$. This method was used by CLEO\cite{CLEO97}, 
resulting in the 
mass distribution shown in Figure~\ref{fig:twocone}.
The sharp drop  near 1.8 
GeV provides the required mass. Combining this result with the BES result for the $\tau$ mass\cite{BES}, CLEO derives an upper limit of 
60 MeV on the neutrino mass at 95\% C.L.

If the tau neutrino were more massive than the tau, the 
tau could not decay: the mass of the neutrino restricts the allowed phase 
space and decay rate.
One can set a limit on the neutrino mass by carefully 
measuring branching ratios of the tau. The theoretically best predicted 
decays are the fully leptonic decays $\tau \to l \nu_l \nu_\tau$ where 
$l  = e, \mu$, with the main uncertainties coming from the tau mass and 
lifetime. These are measured at the per cent level. Knowing the pion 
and kaon form factors, once can also use the single body hadronic decays 
of the tau. Using the Particle Data Group best fit values for all these 
parameters, Swain and Taylor\cite{SWAIN97} derives a 95 \% confidence level 
 upper limit on the tau neutrino mass at 68 MeV.

\section{Accelerator Based Endpoint Limits}
The method that has given the most stringent limits on the tau neutrino 
mass attempts to fit the two dimensional spectrum of hadronic decay product 
mass and energy for $\tau$'s produced at electron-positron colliders. 
\begin{figure}[htb]
\centerline{
\epsfig{file=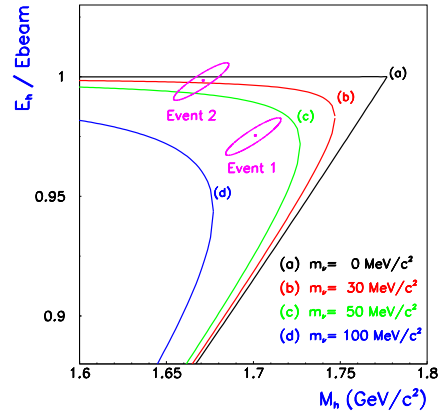,width=60mm,
bbllx=1pt,  bblly=1pt, bburx=555pt, bbury=555pt 
clip=}
}
\caption{A typical scaled hadronic energy vs mass distribution (from 
  \cite{ALEPH98}). Note the kinematically allowed boundaries for different neutrino masses and the typical error ellipses around the events.}
\label{fig:evsm}
\end{figure} 
As shown in Figure~\ref{fig:evsm}, near the endpoint, the kinematically allowed contour 
depends very strongly on the neutrino mass. Not only does the contour 
depend on the neutrino mass, but for any fixed energy, the hadronic mass 
spectrum shape also depends on the neutrino mass: points away from the 
limit contain some information on the neutrino mass. However, for a fixed 
hadronic mass, the derivative of the hadronic energy with respect to the 
neutrino mass is effectively flat except for a delta function at the contour 
limit\footnote{ This means that in the limit of  a detector with no smearing, 
only those points at the contour limit would provide information on the 
neutrino mass in the hadronic energy direction, while along the mass axis 
all points would provide neutrino mass information. 
 The mass limit could 
potentially be set by one point with mis-measured energy - this problem 
is far less severe along the mass axis where all points contain information. 
This approximation points out that the systematic error estimation due 
to smearing has to be done very carefully. There is also a need for a 
careful evaluation of the analytic and statistical properties of the 
likelihood 
function in the presence of an underlying discontinuous estimation parameter 
dependence. No such examination has been done explicitly in any of the 
neutrino mass literature.}.
Finite smearing dilutes further the neutrino 
mass sensitivity, as do initial and final state radiation. Note that ISR/FSR 
cause points to drop along the energy axis in the figure, but do not smear 
them along the mass direction. These effects are calculable however.

One can distinguish two different types of decays used for this kind 
of analysis with different sensitivities to systematic errors. Decays 
to many pions (e.g. $\tau \to 5\pi \nu$)  tend to produce events near the 
sensitive endpoint but suffer from small branching ratios and low reconstruction 
efficiencies.  Decays with few particles in the final state (e.g. $\tau 
\to 3\pi \nu$ ) tend to produce events far away from the endpoint, but 
have efficiencies and branching ratios that are large - one hopes that 
the efficiencies are large enough that a rare event near the endpoint 
will be reconstructed.

The most general likelihood function is of the form:
%
$$
{\cal L}( M_{\nu_\tau} ) = {\cal P }\left( N_{obs} , \mnut \right)
        \prod_{Data}
        (\alpha {\cal L}_{Signal} + (1-\alpha) {\cal L}_{BGD})   
$$

The first term, used only by \cite{CLEO98} and \cite{CLEO2000}, is a Poisson term relating the number of 
 events near the endpoint as a function of the number of events far from the
 endpoint and the neutrino mass. One uses this term as a compromise to
 avoid fitting over many points which do not have much neutrino mass
 information, and also  lessen the  dependence on the explicit form of 
 the physics function in the low mass region. 

The product term is taken over
 all accepted data events, and is composed of a signal term ${\cal L}_{Signal}$ 
 and a background term ${\cal L}_{BGD}$. 
 The signal term is composed of a convolution of the
 expected detector smearing, efficiency and the differential decay
 width. The neutrino mass enters only through the decay width
  and the limits of integration for the convolution
 - again, the energy dependence on neutrino mass
 in this term comes in only as an integration limit, and not a slope. The 
 differential decay width depends on the V-A nature of the weak interaction, 
 the underlying physics of the event and the radiative corrections, including
 ISR and FSR.
 The background term is a parameterization of events which are not signal,
 whether originating from (mis-reconstructed) tau decays or other 
 physics. These terms in general do not depend on the neutrino mass  - this is
 true to a high degree even for the mis-reconstructed $\tau$ events since they tend
 to be highly smeared over the plot. This lack of dependence on neutrino mass
 makes this type of likelihood divergent when all neutrino 
 masses are considered
 so one introduces a prior distribution, cutting off neutrino masses greater
 than say 100 MeV. The background shape is generally determined either from
 Monte Carlo or directly from non-tau decay data.
 The relative weight of signal and background, $\alpha$, is usually 
 determined by fitting for the number of events above the endpoint region. 

 The likelihood is then used to determine an upper limit on the mass
 of the neutrino. This is generally done by integrating\footnote{Note that strictly speaking it is a product of the likelihood
 function and a top hat prior distribution which is being integrated, 
 as per \cite{PDG}} the likelihood
 to a 95th percentile above 0 MeV
 or by finding where the log of the likelihood
 drops by 1.92 from its peak. Neither method is more correct than the other, 
 rather each expresses a different philosophy on the meaning of the upper 
 limit. In all cases published so far the application of either method gives
 a consistent answer to  within $\approx{\cal O}(5 {\rm MeV}) $.

\section{ Current Results }
\begin{figure}[htb]
\centerline{
\epsfig{file=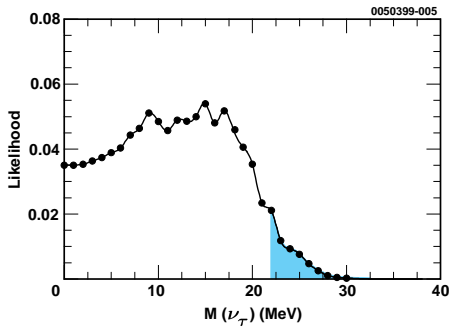,width=60mm ,clip=}
}
\caption{ The neutrino mass likelihood distribution for CLEO's $\tau 
\to 3\pi\pi^0\nu$ sample. Note the raw 95th percentile  at 22 MeV. }
\label{fig:likelicleo2000}
\end{figure}

Since only one new result has been published since the last conference, readers
 desiring a full review of previous results should refer to R. McNulty's review
 in the previous set of proceedings \cite{McNulty}. These results are briefly summarized in
Table~\ref{table:Summary}. The new result is
from  CLEO\cite{CLEO2000} 
 which
has  examined  the 
  neutrino mass using the decay $\tau \to 3\pi \pi^0 \nu$ for the first time.
\begin{figure}[htb]
\centerline{
\epsfig{file=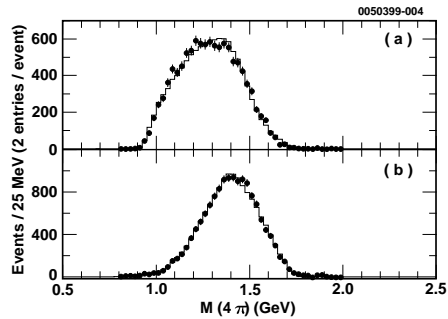,width=60mm ,clip=}
}
\caption{Four pion invariant masses for events with an $\omega$ (top) 
 and events
with out an $\omega$ (bottom) for CLEO's $\tau \to 3\pi\pi^0\nu$  sample.
 The histogram shows the Monte Carlo expectation. }
\label{fig:4pimasscleo2000}
\end{figure} 
  This sample is
 composed of 29,000 decay events, as shown in  
Figure~\ref{fig:4pimasscleo2000}.
 The fit includes an explicit 
 term accounting function for mis-identified tau decays, corresponding to 7\% of
 the event sample. Some 3\% of the sample is accounted for by a non-tau 
 background  term. The spectral
 function in this decay was fit below the neutrino mass sensitive region
to a combination of non-resonant, rho and omega subcomponents. This fit was
 then extended into the endpoint region. 
 The resulting likelihood distribution is shown in Figure~\ref{fig:likelicleo2000}.
 The 95th percentile  above 0 MeV is at 22 MeV for the raw data, and
 the total systematic error, added in linearly to 
 conform with the standard set by
 previous studies, sets the final limit at 28 MeV. The main systematic
 error comes in through the spectral function dependence, the $\pi^0$ energy
 scale, as well as the charged track momentum reconstruction.

\section{ General Remarks on Published Results }
A glance at Table~\ref{table:Summary}  summarizing all published results shows that
 all limits are generally within the same range, in spite of a very 
large range of event sample sizes.

\begin{sidewaystable}
\caption{Summary of Neutrino Mass Limits }
\label{table:Summary}
\newcommand{\m}{\hphantom{$-$}}
\newcommand{\cc}[1]{\multicolumn{1}{c}{#1}}
\renewcommand{\tabcolsep}{2pc} 
\renewcommand{\arraystretch}{1.2} 
\begin{tabular}{@{}llllll}
\hline
	 & Decay Mode	& Events & $< \sigma_M >$ (MeV) & $M_{95}$ (MeV)  & 
$\approx \hat{m}  (MeV)$ \\
\hline	
ALEPH98\cite{ALEPH98}	 & $5\pi(\pi^0)$       & \m55 			& \m15  	& 
\m23 &  0  \\
CLEO98\cite{CLEO98}   & $5\pi,3\pi2\pi^0$ 	& \m$\approx 450$ 	& \m15 		& 
\m30 & 15 \\
OPAL98\cite{OPAL98}   & $5\pi$ 	& \m22			& \m25 (?)	& \m43.2
& 0 \\
ALEPH95\cite{ALEPH95}  & $5\pi$ 	& \m25			& \m15		& \m24 &
 0  \\
OPAL95\cite{OPAL95}	 & $5\pi$	& \m5			& \m25		& \m74 &
 0  \\
CLEO93\cite{CLEO93}	 & $5\pi,3\pi2\pi^0$	& \m113			& \m10		& \m32.6  & 0 \\
ARGUS92\cite{ARGUS92}	 & $5\pi$	& \m19			& \m10 (?)	& \m31 & (*) \\
CLEO00\cite{CLEO2000}	 & $3\pi\pi^0$  & \m$\approx 17,000$	& \m20		& \m28  &
 0 \\
ALEPH98\cite{ALEPH98}	 & $3\pi$	& \m$\approx 2,900$	& \m15		& \m22 & 
 5 \\
OPAL97\cite{OPAL97}	 & $3\pi$	& \m$\approx 2500$	& 
			$\sigma_{M_{miss}^2} \approx 0.1 GeV^2$ &
			\m29.9  & 0 \\
DELPHI97\cite{McNulty} & $3\pi$	& \m$\approx 12,000$ 	& \m20		& \m28(68) & -10 \\
\hline
\end{tabular}\\[2pt]
The $< \sigma_M >$ gives the typical mass smearing for the reconstructed event
masses.
The $\hat{M}$ column denotes the neutrino mass value which maximizes likelihood as
 estimated by this author from published likelihoods.\\
OPAL97\cite{OPAL97} is a two dimensional fit of missing mass vs missing energy.
DELPHI97\cite{McNulty} notes that its limit depends strongly on the admixture of a 
 possible high mass $3\pi$ state, with a 28 MeV limit for no admixture, and
 a 68 MeV for a large admixture.
ARGUS92\cite{ARGUS92} did not publish a likelihood curve.
\end{sidewaystable}

The typical upper limit in this table is in the 30 MeV range. One would
 naively expect that the scatter in the most likely neutrino mass for each
 experiment should be on the order of $30 MeV /1.64 = 18 MeV$ - this is
 manifestly not the case. For the purposes of the following, let us 
 imagine that the true neutrino mass is 0 MeV, and that each experiment
 has a resolution of 18 MeV on the neutrino mass. I believe that 7 of the
  published results in the table are statistically independent, in addition
 to the unpublished DELPHI result\cite{McNulty}.
 Once can  ask what the Gaussian probability 
 is that all 7 different published experiments 
  could all arrive at the result that 
 the most likely neutrino mass is between 0 MeV and 18 MeV - 
 this is obtained from a one tailed Gaussian distribution and is 
 equal to $0.64^{7} = 4 \% $. If we include the unpublished 
 DELPHI97\cite{McNulty} result and ask what the probability is that the most likely
 mass is less than 18 MeV, including negative values, the probability
 is somewhat more promising, namely 
$ { 8 ! \over{ 7! 1!}} * 0.84^{7} *0.16^{1} = 38 \% $. 
 In this case however, the probability of all the maxima being
 located within $\pm 18 MeV $ of 0 MeV is derived from a two-tailed 
 Gaussian as $0.68^{8} = 5\%.$ If one argues that in fact the typical
 resolution is larger than 18 MeV, then these results are even
 more extreme. 
 These simple considerations lead me to
 conclude that some subtle series of systematic biases must be driving
 down the published neutrino mass limit\footnote{ One might also argue that
  the Gaussian 
 approximation used in the above is invalid for such small statistics - 
 this of course calls into question the meaning and validity of the
 quoted upper limits.}.

\section{Subtleties}

 How should one quote the neutrino mass limit?
 There is no accepted way to judge the relative 
 merits of different limits.  The  PDG\cite{PDG}
 simply quotes the lowest published value for the
 upper limit. This prescription is unfortunately biased towards finding
 a massless neutrino. 
 
One can attempt to combine the likelihood
 functions of different experiments and thus hope to obtain a 
 clear answer. However, should any one of those experiments suffer from
 an unfortunate fluctuation due to an undetected systematic error towards 
 low $m_\nu$, the
 resulting combined distribution will also suffer from this. 
 The problem with this method 
 is that 
 all of the experimental results are considered on an equal footing. 
 One should clearly
 have a mechanism to de-weight less reliable results. Combining limits 
 from different published results is simply
 quite difficult. Not only is it not clear how to incorporate different 
 systematic  errors from different experiments properly, but the fact 
 that the different  measurements tend to use slightly different 
 functional forms for their likelihood makes a blind multiplication of
 likelihood functions difficult to interpret.  

 One might argue that,
 in the face of similar experimental resolutions, the experiment with
 the largest sample should have the most robust result, just like in
 the case of a branching ratio measurement.
 Neutrino mass limits differ in some important ways from the limits
 one typically makes on branching ratios. A branching ratio limit can 
 treat each event as having an equal amount
 of information. However, the information content of an event in a neutrino
 mass limit depends on its reconstructed location in the energy vs mass plane 
 and on its associated expected smearing, as well as the particular spectral 
 function of the mode under consideration. This means that the large N limit
 where we are all typically comfortable in interpreting likelihoods as
 statistically meaningful depends on what region of neutrino mass one is
 investigating. ``Large N'' for a 100 MeV limit can be much smaller than 
 ``large N'' for a 10 MeV limit\footnote{ Similarly ``large N'' 
 for a mode such as $\tau \to 3\pi\nu$
 has to be much larger than ``Large N'' for the mode $\tau \to 5\pi\nu$.}.
 In addition, the range of possible  reconstructed
 masses for an event depends on the value of the true neutrino mass 
 (this is most clearly seen for a perfect detector with delta
  function resolution.) An important property of
 maximum likelihood estimators is that they are efficient: 
 if an estimate of minimum variance
 exists for a parameter, then the maximum likelihood should 
 find it\cite{PDG}.
 However, the derivation of this minimum variance bound,  known as 
 the Rao-Cramer-Frechet bound, 
 requires that  the range of the underlying event probability
  distribution not depend
 on the parameter being estimated\cite{PDG}: this is explicitly not true for
 the neutrino mass fits. 
  Thus, with the
 Rao-Cramer-Frechet bound in question, it is not clear that the maximum 
 likelihood estimator can find the estimate with the minimum 
variance\footnote{  One might argue that the presence of smearing, which 
 allows events to be reconstructed above the kinematically determined endpoint,
  makes the above argument  less compelling. However,
this line of thought leads to the conclusion that a perfect detector
 with no smearing is less useful for determining a neutrino mass than
 one with smearing. Such a repugnant thought is not consistent with
 our understanding of how measurements should work.}.
 Of course in the opposite extreme, if the smearing
 is large enough, the information content of any point tends to zero, 
 and the minimum  variance bound tends to infinity. These observations point 
 out the need  
 for a serious evaluation of the form of the likelihood used for neutrino
 mass estimation and  its properties.

 One should also keep in mind that showing consistency with a massless
 neutrino as one typically does by quoting an upper limit is not the
 same as excluding massive neutrinos. One can occasionally get low fluctuations
 in the neutrino mass limit even for massive neutrinos.\footnote{ See 
 hypothesis testing, and Type I and Type II errors in any introductory 
 statistics  book.}

 Of course one very important consideration that must be taken into account
 is the sociological bias towards massless neutrino results. Experimental
 groups tend not to work as hard to publish results that are not more stringent
 that their competitor's. 

\section{Upper Limits Are Not What You Think They Are}
 As an illustration that neutrino mass limits are not straightforward, consider
 Table~\ref{table:Oddity}. This table summarizes the number of times one
 gets an upper limit at 27 MeV or lower using the CLEO Monte Carlo for 
 reconstructed data samples of 25 events and 450 events for two different
 input neutrino masses of 0 MeV and 50 MeV. An experimenter is confronted with
  an experimentally derived limit, and from this must venture a guess as to
 what the input (true) neutrino mass might have been. With this in mind,
 with a small sample of 25 events one is much more likely to mistakenly
 accept or reject a massive neutrino than one is with a large 450 event sample.
 Thus with similar detectors at similar energies and the same decay mode,
  one is more likely to
 infer the correct conclusion about the neutrino mass using  
 information from the largest reconstructed sample 
 instead of using the lowest estimated upper limit. It is statistics,
 not lucky events near the endpoint that give discriminatory power. It is also
 interesting that even with a sample as large as 450 events, the figure of
 $95\%$ appears nowhere in this table - this indicates that the variance in
 results is still not small.

\begin{table*}[htb]
\caption{The probability of obtaining a 95th percentile
  mass limit of 27 MeV or lower
 for different input neutrino masses and event sample sizes.}
\label{table:Oddity}
\newcommand{\m}{\hphantom{$-$}}
\newcommand{\cc}[1]{\multicolumn{1}{c}{#1}}
\renewcommand{\tabcolsep}{2pc} 
\renewcommand{\arraystretch}{1.2} 
\begin{tabular}{@{}lll}
\hline
$M_\nu^{input}   $   & 25 Events  & 450 Events  \\
\hline	
$ 0 {\,\rm MeV } $     	    & \m3\%      & \m67\%      \\
$ 50 {\,\rm MeV} $           & \m1\%      & \m$\ll$ 1\%   \\
\hline
\end{tabular}\\[2pt]
The values here are calculated using the CLEO Monte Carlo and detector simulation for $\tau \to 5\pi\nu$ decays. Exact values will vary from experiment 
 to experiment, and decay mode to decay mode.
\end{table*}

\section{ Exhortations to future limit setters}

 In light of the observations made above, I would like to counsel future
 limit setters to heed the following advice:
\begin{itemize}
\item Restrict your fit region and use a Poisson Term: Since most events
 far from the endpoint have little neutrino mass information apart
 from helping to normalize the expected number of events near the 
 endpoint, it is not worth CPU time to fit these events in detail. 
\item Use a Background Function: If one wishes to avoid using a 
 background function, one must use tighter and tighter cuts as 
 the accepted luminosity increases. This can result in the paradoxical
 situation in which an experiment ends up with less sensitivity as its
 sample size grows. The inclusion of a background function remedies this
 strange state of affairs.
\item Publish the expected reach $<M_{95}>$ (and its variance) for
 a massless neutrino hypothesis. This will help establish whether
 the limit from data is meaningful or a dangerous fluctuation.
\item Publish the discriminatory power of your experiment: how often
 does your data derived limit occur for a large (say 50 MeV) neutrino
 mass?
\item Compare the  limit derived from mass information only to the 
  limit derived from the energy vs mass distribution: 
  if the two dimensional limit is substantially 
 different from the one dimensional one, then the limit is being driven
  by a small number  of events or a very unlikely event distribution.
\item Carefully examine large smearing tails: comments given in the text
 point out that the sharp cut-off in reconstructed energy 
  as determined by the neutrino mass make the properties of the
 neutrino mass estimator unclear. At the very least, one should 
 carefully examine the effects of systematic error along the energy axis.
 Ultimately, as samples grow and the relevant fit region shrinks, fluctuations
 due to the unknown tails of the smearing functions  in both mass and energy 
 must dominate the fits.
\end{itemize}

\section{ How to Set a Limit We Can All Understand }

 In light of the suspicious properties of the maximum likelihood
 estimator raised in this review, it would be very useful if experimenters
 would calibrate their estimators using a Monte Carlo method. The method
  proposed herein answers the question of ``What masses are ruled out by
 my data ?''
\begin{itemize}
\item Generate a large number of Monte Carlo samples with 
 sizes comparable to those obtained in data for different 
 input neutrino masses  ( $m_\nu^{in}$) 
\item For each of these samples, form the likelihood, and 
 find the neutrino mass which 
 maximizes the likelihood, $\hat{m}_\nu^{MC}$
\item Plot $\hat{m}_\nu^{MC}$ vs $m_\nu^{in}$ 
\item Form the likelihood for the data sample and find its maximum 
$\hat{m}^{Data}$
\item Find $m_\nu^{in}$ such that  $\hat{m}_\nu^{MC}$ is larger than
 $\hat{m}^{Data}$ $95\%$ of the time
\item Set the upper limit to this value
\end{itemize}

Unfortunately this method does not provide a nice way to combine
 limits from different experiments. However given the points made in
 this review, perhaps it is better to suffer from this than to combine
 likelihoods with unclear statistical properties, and possibly 
 incompatible meanings.

\section{Conclusions}
 I have briefly reviewed the current constraints on an MeV scale tau neutrino
 mass. These constraints are from astrophysical observations as well as from
 terrestrial observations  and all of them taken at face value
 allow for the existence of an unstable $\nu_\tau$ with a mass on an
 MeV scale. Recent SuperKamiokande results indicate that the (stable) neutrinos
 participating in atmospheric $\nu_\mu$  disappearance have a mass well below
 an MeV and are consistent with stable tau neutrinos. 
 An inspection of the two dimensional
 limits on the $\tau$ neutrino mass derived from experimental data reveal
 that the ensemble of limits is too consistent. I have also for the first
 time shown that the usual assumption that the likelihood is well behaved for
 this technique is questionable. I have ended by recommending what extra
 information future limit setters should publish to allow others to gauge
 the believability of their limits, and have also recommended a method
 more suited to obtaining an answer to the question of what neutrino
 mass is excluded by the observed data than the methods used up to now.

\section{Acknowledgments}
 The audience would have found this talk
 far more tedious had it
 not been for many 
 interesting discussions with Ronan McNulty and many other colleagues at 
 CERN and on CLEO.
 I'd like to thank
 the conference organizers for inviting me to give this
 talk in such a nice location. 
 I also would like to thank the National Science
 Foundation for its support.


\begin{thebibliography}{9}
\bibitem{Zeldovich66} S.S.~Gershtein. and Ya.B.~Zel'dovich,  JETP Lett. {\bf 4} (1966) 174.
\bibitem{Cowsik} R.~Cowsik and J.~McClelland, Ap.J. {\bf 180} (1973) 7.
\bibitem{Kolb} E.~Kolb and M.~Turner, The Early Universe, Addison-Wesley, 1990.
\bibitem{Turner00} M. Turner, private communication.
\bibitem{ALEPH98}  B.~Barate et~al., ALEPH coll., Euro. Phys. J. {\bf C2} 
(1998) 395.
\bibitem{KARMEN} B.~Armbruster et~al., KARMEN coll., Phys. Lett. {\bf B348} 
(1999) 19.
\bibitem{CLEO97} A.~Anastassov et~al., CLEO coll., Phys. Rev. {\bf D55} (1997) 2559.
\bibitem{BES} J.Z. Bai et~al., BES Collaboration, Phys. Rev. {\bf D53} (1996) 20.
\bibitem{SWAIN97} J.~Swain and L.~Taylor, Phys. Rev. {\bf D55} (1997) 1.
\bibitem{CLEO98} R.~Ammar et~al., CLEO coll., Phys. Lett. {\bf B431} (1998)
 209.
\bibitem{McNulty} R.~McNulty, Nuclear Physics B (Proc. Suppl.) 76 (1999) 409.
\bibitem{CLEO2000} M.~Athanas et~al., Phys. Rev. {\bf D61} (2000) 052002.
\bibitem{PDG} The Particle Data Group, Euro. Phys. J. {\bf C15} 2000.
\bibitem{OPAL98} K.~Ackerstaff et~al., OPAL coll., Euro. Phys. J. {\bf C5} 
(1998) 229.
\bibitem{ALEPH95} D.~Buskulic et~al., Phys. Lett. {\bf B349} (1995) 585.
\bibitem{OPAL95} R.~Akers et~al., Z. Phys.  {\bf C65} (1995) 183.
\bibitem{CLEO93} D.~Cinabro et~al., Phys. Rev. Lett. {\bf 70} (1993) 3700.
\bibitem{ARGUS92} H.~Schroder et~al., Mod. Phys. Lett. {\bf A8} (1993) 573.
\bibitem{OPAL97} G.~Alexander et~al., OPAL coll., Z. Phys. {\bf C72} (1996)
 231.
\end{thebibliography}
\end{document}